\documentclass[prd,tightenlines,nofootinbib,prd]{revtex4}
\usepackage{amsmath,amscd,amssymb}
\usepackage{color,epsfig}
\usepackage{xfrac}
\usepackage{latexsym}
\usepackage{mathrsfs}
\usepackage{graphicx}
\usepackage{bbm}      
\usepackage{diagbox}

\newcommand{\imu}{{\rm i}}

\newcommand{\zr}[1]{\mbox{\hspace*{#1em}}}
\newcommand{\ID}{\mbox{{\sf 1}\zr{-0.16}\rule{0.04em}{1.55ex}\zr{0.1}}}
\newcommand{\fract}[2]{{\textstyle\frac{#1}{#2}}}

\usepackage[normalem]{ulem}

\usepackage[normalem]{ulem}

\begin{document}

\title{Vacuum polarization energy of the Shifman-Voloshin soliton}

\author{H. Weigel$^{a)}$, N. Graham$^{b)}$}

\affiliation{
$^{a)}$Institute for Theoretical Physics, Physics Department, 
Stellenbosch University, Matieland 7602, South Africa\\
$^{b)}$Department of Physics, Middlebury College
Middlebury, VT 05753, USA}

\begin{abstract}
We compute the vacuum polarization energy of soliton configurations in a model with two 
scalar fields in one space dimension using spectral methods. The second field 
represents an extension of the conventional $\phi^4$ kink soliton model. We find that 
the vacuum polarization energy destabilizes the soliton except when the fields have 
identical masses. In that case the model is equivalent to two independent $\phi^4$ models.
\end{abstract}

\maketitle

\section{Introduction}

Two-field models supporting solitons in one space dimension obtainable as 
Bogomol'nyi-Prasad-Sommerfeld (BPS) solutions have been considered in the 
context of a number of applications, including supersymmetry and domain 
walls, see~\cite{Abraham:1990nz,Cvetic:1991vp,Cecotti:1992rm,Bazeia:1995en,Chibisov:1997rc,Shifman:1997wg,Bazeia:2004dh} 
and references therein. The essential feature leading to these applications 
is that in one space dimension the soliton has a localized kink shape, which 
becomes a surface (domain wall) when embedded in higher dimensions. When two (or
more) fields interact multiple kinks at finite separation(s) emerge. The BPS 
construction is then carried out by writing a superpotential for the fields.
The simplest such model has been introduced by Bazeia {\it et al.}~\cite{Bazeia:1995en} who 
also constructed some of its soliton solutions, while the full spectrum of solitons, 
including numerical simulations, was uncovered by Shifman and Voloshin~\cite{Shifman:1997wg}. 
In Ref.~\cite{Weigel:2017kgy} the analytically known solitons of this model were considered as 
an illustration of general techniques allowing for the extension of scattering theory methods 
for computing one-loop quantum corrections \cite{Graham:2009zz} to the case of models with a 
mass gap, which then have multiple thresholds in the scattering problem. Such corrections were
computed in that model for the simple cases where the soliton does not couple the fluctuation 
modes of the two fields in Ref.~\cite{AlonsoIzquierdo:2012tw} and for small and moderate
separation of the kinks in Refs. \cite{AlonsoIzquierdo:2003gh,Alonso-Izquierdo:2013pma}  
using heat kernel methods~\cite{Elizalde:1996zk}.

In this Letter, we apply the methods of Ref.~\cite{Weigel:2017kgy} to study these quantum 
corrections in more detail by going beyond the analytically known solitons for this particular 
model, which we define following the approach and conventions of Ref.~\cite{Bazeia:1995en}. 
We show that quantum corrections can significantly alter the classical stability of solitons 
in this model. In particular, the model can become unstable to the formation of a kink-antikink 
pair separated by a large region of {\it secondary} vacuum, whose classical energy density 
equals that of the {\it primary} vacuum outside the kink-antikink pair, but whose one-loop 
quantum energy density is negative.

The Bazeia model extends the $\phi^4$ model by a second scalar field $\chi$. Its Lagrangian reads
\begin{equation}
\mathcal{L}=\frac{1}{2}\left[\partial_\nu \phi\partial^\nu \phi
+\partial_\nu \chi\partial^\nu \chi\right]
-\frac{\lambda}{4}\left[\phi^2-\frac{M^2}{2\lambda}
+\frac{\mu}{2}\chi^2\right]^2-\frac{\lambda}{4}\mu^2\chi^2\phi^2\,.
\label{eq:Bazeia}\end{equation}
The Lagrangian contains the typical coupling constant $\lambda$ and the mass scale $M$ as in 
the conventional $\phi^4$ model. We will discuss the meaning of the dimensionless coupling 
constant $\mu$ shortly. First we note that in the case $\mu=2$, when the two fields are
indistinguishable,  the orthogonal transformation
$\varphi_{1,2}=\frac{1}{\sqrt{2}}\left[\chi\pm\phi\right]$ decouples the model into
$$
\left[\phi^2-\frac{M^2}{2\lambda}+\chi^2\right]^2+\lambda\chi^2\phi^2
=2\left[\varphi_1^2-\frac{M^2}{4\lambda}\right]^2+
2\left[\varphi_2^2-\frac{M^2}{4\lambda}\right]^2\,,
$$
which is a sum of two conventional and identical $\phi^4$ models. As a result, the 
known results \cite{Ra82} from the $\phi^4$ model with its kink soliton solution will 
provide checks of our calculations.

There are two distinct vacuum configurations. First, the solution with 
$\phi=\pm M/\sqrt{2\lambda}$ and $\chi=0$, adopted from the  $\phi^4$ model,
and second, the solution with $\phi=0$ and $\chi=\pm M/\sqrt{\mu\lambda}$. Later 
we will see that only the first allows for BPS soliton solutions unless $\mu=2$, 
and thus we refer to it as the {\it primary} vacuum and the second as the 
{\it secondary} vacuum. The masses for fluctuations of the primary vacuum are 
$m_\phi=M$ and $m_\chi=\mu M/2$. That is, the dimensionless coupling constant 
is twice the ratio of the two masses.

After appropriate redefinition of the fields, 
$(\phi,\chi)\to(M/\sqrt{2\lambda})(\phi,\chi)$ and the 
coordinates, $x_\nu\to2x_\nu/M$ the rescaled Lagrangian,
$\mathcal{L}\to(M^4/8\lambda)\mathcal{L}$
is conveniently expressed as
\begin{equation}
\mathcal{L}=\frac{1}{2}\left[\partial_\nu \phi\partial^\nu \phi
+\partial_\nu \chi\partial^\nu \chi\right]-U(\phi,\chi)
\qquad {\rm with}\qquad
U(\phi,\chi)=\frac{1}{2}\left[\phi^2-1+\frac{\mu}{2}\chi^2\right]^2
+\frac{\mu^2}{2}\phi^2\chi^2\,.
\label{eq:fpot} \end{equation}
In these units the primary vacuum configuration is $\phi_{\rm vac}=\pm1$ and $\chi_{\rm vac}=0$ 
so that $m_\chi=\mu$ and $m_\phi=2$. Note that with these dimensionless variables the classical 
mass is measured in units of $M^3/\lambda$, while the one-loop quantum energy, which is central 
to the current study, scales with $M/m_\phi$. The different scales arise from the overall 
loop-counting factor in $\mathcal{L}$ that emerges from canonical quantization.

In Section 2 we describe the construction of the solitons in this model.
Following, in Section 3, we review the computation of the one-loop quantum,
or vacuum polarization energy (VPE) in the no-tadpole renormalization scheme. 
In Section 4 we present the numerical results for the VPE and show that it produces 
an instability unless $\mu=2$. We conclude in Section 5. In an Appendix we show 
that the finite renormalization imposing on-shell conditions does not alter the 
conclusion of instability.

\section{Soliton}

The Bazeia model \cite{Bazeia:1995en} is defined to allow a 
BPS construction for the classical energy
\begin{align}
E_{\rm cl}&=\frac{1}{2}\int_{-\infty}^\infty dx\,
\left[\phi^{\prime2}+\chi^{\prime2} +\left(\phi^2-1+\frac{\mu}{2}\chi^2\right)^2
+\mu^2\phi^2\chi^2\right]\cr
&=\frac{1}{2}\int_{-\infty}^\infty dx\,
\left[\left(\phi^2-1+\frac{\mu}{2}\chi^2\pm\phi^\prime\right)^2
+\left(\mu\phi\chi\pm\chi^\prime\right)^2\right]
\pm\left[\phi-\frac{1}{3}\phi^3-\mu\phi\chi^2\right]_{-\infty}^\infty\,,
\label{eq:bps}\end{align}
where the prime denotes the derivative with respect to the (dimensionless) 
coordinate $x$. We immediately see that only profile functions that assume
the primary vacuum configuration can have finite non-zero energy.\footnote{For 
$\mu=2$ an alternative BPS construction is possible producing a soliton with
$\lim_{|x|\to\infty}\chi(x)\ne0$.} Choosing $\phi(\pm\infty)=\pm1$ requires 
the upper sign in Eq.~(\ref{eq:bps}) because $\phi(x)$ must (monotonically) 
increase. Then the BPS equations read
\begin{equation}
\frac{d\chi(x)}{dx}=-\mu\phi(x)\chi(x)
\qquad{\rm and}\qquad
\frac{d\phi(x)}{dx}=1-\phi^2(x)-\frac{\mu}{2}\chi^2(x)\,.
\label{eq:lindeq}\end{equation}
These coupled differential equations have been studied in detail by Shifman and 
Voloshin \cite{Shifman:1997wg}. For completeness we discuss those results.
The model exhibits translational invariance and we center the (eventual)
soliton at $x_0=0$. Then $\chi$ and $\phi$ are symmetric and anti-symmetric
functions, respectively,\footnote{Eqs.~(\ref{eq:lindeq}) also allow the opposite
choice; but then the energy, Eq.~(\ref{eq:bps}) is zero.} and so $\phi(0)=0$ and 
$\chi^\prime(0)=0$. We are free to choose $\chi(0)\ge0$. If $\chi(0)>\sqrt{2/\mu}$, 
$\phi^\prime(0)<0$ so that $\phi(0^+)<0$. In turn $\chi$ would increase 
and $\chi(0)$ would be a minimum. Furthermore $\phi^\prime$ would turn 
even more negative and not approach $+1$ at spatial infinity. By contradiction 
we thus conclude that $\sqrt{2/\mu}$ is an upper bound for $\chi(0)$ 
and we parameterize $\chi(0)=a\sqrt{2/\mu}$ with $0\le a<1$. An equivalent 
bound was derived in Ref.~\cite{Shifman:1997wg} from the condition
that the solution to
$$
\frac{d\phi^2}{d\chi}=2\phi\frac{d\phi}{dx}\left(\frac{d\chi}{dx}\right)^{-1}
=-\frac{2-2\phi^2-\mu\chi^2}{\mu\chi}
$$
is consistent with $\phi^2\ge0$.

Because of the reflection symmetry $x\leftrightarrow-x$ it is sufficient 
to solve Eqs.~(\ref{eq:lindeq}) on the half-line $x\ge0$. In the numerical 
simulation we initialize $\phi(0)=0$ and $\chi(0)=a\sqrt{2/\mu}$ and
vary $a$. For any numerical solution we then verify that the first integral in 
Eq.~(\ref{eq:bps}) produces $E_{\rm cl}=\frac{4}{3}$. We also verify that the 
numerical solutions agree with the analytically known results listed in 
Table~\ref{tab:known}.  

\renewcommand{\arraystretch}{1.3}
\begin{table}[htbp]
\centerline{
\begin{tabular}{c|c|c|c}
& $\phi(x)$ & $\chi(x)$ & parameters\cr
\hline
I) & ${\rm tanh}(x)$     &    $0$     & $a=0$ \cr
II) & ${\rm tanh}(\mu x)$ & $\frac{\sqrt{2\left(1/\mu-1\right)}}{{\rm cosh}(\mu x)}$ &
$\mu<1\,,$~~$a=\sqrt{1-\mu}$ \cr
III) & $\frac{{\rm sinh}(2x)}{b+{\rm cosh}(2x)} $ &
$\frac{\sqrt{b^2-1}}{b+{\rm cosh}(2x)}$ & $\mu=2\,,$~~$b=\frac{1+a^2}{1-a^2}$\cr
IV) & $\frac{(1-a^2){\rm sinh}(x)}{a^2+(1-a^2){\rm cosh}(x)}$ &
$\frac{2a}{\sqrt{a^2+(1-a^2){\rm cosh}(x)}}$ & $\mu=\frac{1}{2}$
\end{tabular}}
\caption{\label{tab:known}Analytically known soliton solutions \cite{Bazeia:1995en,Shifman:1997wg}.}
\end{table}
\renewcommand{\arraystretch}{1.0}
We thus find that the various known analytical solutions are not independent but are related 
by a single parameter. If these solitons were independent, a third zero mode for the small
amplitude fluctuations about the soliton along the direction in field space connecting the 
solutions would have emerged, but only two have been observed \cite{Weigel:2017kgy}. Stated 
otherwise, the solitons are parameterized by two continuous parameters \cite{Shifman:1997wg}: 
the center of the soliton, which we set to zero, and the amplitude of the $\chi$ field, which 
we parameterize by $a$. Varying these parameters produces the two observed zero modes. 
Alternatively, the family of solitons can be constructed by successively adding infinitesimal 
contributions proportional to the zero mode wave-function.

\begin{figure}
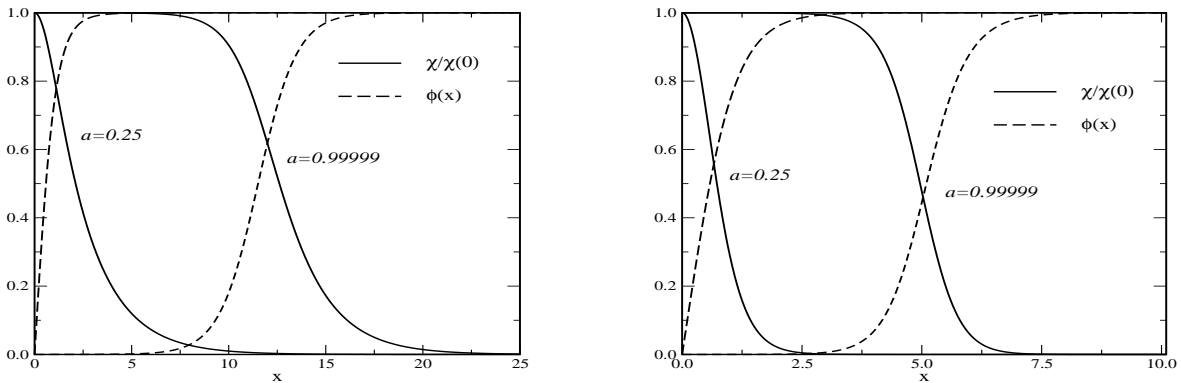

\centerline{
\epsfig{file=sol05.eps,height=5cm,width=7cm}\hspace{1.5cm}
\epsfig{file=sol30.eps,height=5cm,width=7cm}}
\caption{\label{fig:sol}Soliton profiles for $\mu=0.5$ (left panel) and 
$\mu=3.0$ (right panel). Note the different scales on the horizontal
axes.}
\end{figure}

The limit $a\to1$ deserves further discussion. In that case, the right-hand-sides of 
Eq.~(\ref{eq:lindeq}) are tiny in a wide region around $x=0$, so that the profiles stay
constant at their $x=0$ values. Eventually two well separated structures
emerge at which $\phi$ changes from $-1$ to zero and zero to $+1$, 
respectively \cite{Shifman:1997wg}. Simultaneously, $\chi$ changes from zero to 
$a\sqrt{2/\mu}$ and back to zero. We show this
behavior in Fig. \ref{fig:sol} (where we only display the
$x\ge0$ regime since the profiles are obtained by
reflection for $x\le0$).
When $a\to1$, $\chi(0)$
approaches $\sqrt{2/\mu}$ and the slope $\phi^\prime(0)$ decreases so
that the profiles assume the secondary vacuum configuration in a large range of coordinate space.

While $\mu$ is a model parameter, $a$ is a variational parameter that we tune 
to minimize the total energy. Since $E_{\rm cl}$ does not depend on $a$, we only 
need to consider the $a$ dependence of the VPE, whose formulation we discuss next.

\section{Vacuum Polarization Energy}

The computation of the vacuum polarization energy in models with a mass gap ($\mu\ne2$)
has been established in Ref. \cite{Weigel:2017kgy}. We briefly summarize it here. The 
central input is the Jost function for imaginary momenta $t=\imu k$. The starting point 
for its computation is the second order differential equation
\begin{equation}
Z^{\prime\prime}(t,x)=2Z^\prime(t,x)D(t)+M^2Z(t,x)-Z(t,x)M^2+V(x)Z(t,x)
\qquad {\rm with}\qquad 
M^2=\begin{pmatrix} \mu^2 & 0 \cr 0 & 4\end{pmatrix}
\label{eq:master} \end{equation}
and
\begin{equation}
D(t)=\begin{pmatrix} t_1 & 0 \cr 0 & t_2\end{pmatrix}\,,
\qquad{\rm where}\qquad
\left(t_1\,,t_2\right)=\begin{cases}
\left(t\,,\sqrt{t^2-\mu^2+4}\right) & \quad{\rm for}\quad \mu\le 2\\[3mm]
\left(\sqrt{t^2-4+\mu^2}\,,t\right) & \quad{\rm for}\quad \mu\ge 2\,.
\end{cases}
\label{eq:deft1t2}\end{equation}
The scattering problem is defined via the potential matrix
\begin{equation}
V(x) \equiv 
\partial^2 U - M^2 =\begin{pmatrix}
\mu(1+\mu)\left(\phi^2-1\right)+\frac{3}{2}\mu^2\chi^2 &
2\mu(1+\mu)\chi\phi \\[2mm] 2\mu(1+\mu)\chi\phi &
6\phi^2-6+\mu(\mu+1)\chi^2
\end{pmatrix}\,,
\label{eq:vpot}
\end{equation}
with the soliton profiles $\phi=\phi(x)$ and $\chi=\chi(x)$ obtained as
(numerical) solutions to Eq.~(\ref{eq:lindeq}) substituted. We solve the matrix 
equations~(\ref{eq:master}) with the boundary conditions
$\lim_{x\to\infty}Z(t,x)=\ID$ and $\lim_{x\to\infty}Z^\prime(t,x)=0$
and extract the Jost matrices
\begin{equation}
F_S(t)=\lim_{x\to0}\left[Z(t,x)-Z^\prime(t,x)D^{-1}(t)\right]
\qquad {\rm and}\qquad
F_A(t)=\lim_{x\to0}Z(t,x)\,.
\label{eq:defjostsym}
\end{equation}
The reflection properties of the profiles lead to skewed parity,
$V(-x)=\begin{pmatrix}1 & 0 \cr 0 & -1\end{pmatrix}\, V(x)\, 
\begin{pmatrix}1 & 0 \cr 0 & -1\end{pmatrix}$ so that the decoupled Jost matrices are
\begin{equation}
F_{\pm}(t)=\left[P_{\pm}F_S(t)D_{\mp}(t)+P_{\mp}F_A(t)D_{\pm}^{-1}(t)\right]\,,
\label{eq:skjost}\end{equation}
with projectors 
$P_{+}=\begin{pmatrix}1 & 0 \cr 0 & 0\end{pmatrix}$,
$P_{-}=\begin{pmatrix}0 & 0 \cr 0 & 1\end{pmatrix}$
and the factor matrices
$$
D_{+}(t)=\begin{pmatrix}-t_1 & 0 \cr 0 & 1\end{pmatrix}
\qquad {\rm and}\qquad
D_{-}(t)=\begin{pmatrix}1 & 0 \cr 0 & -t_2\end{pmatrix}\,.
$$
From this we finally compute the appropriate Jost function for 
imaginary momenta
\begin{equation}
\nu(t) \equiv {\rm ln}\,{\rm det}\left[F_{+}(t) \,F_{-}(t)\right]
\label{eq:defnu}
\end{equation}
that enters the vacuum polarization energy
\begin{equation}
E_{\rm vac} \equiv 
\int_{m_0}^{\infty} \frac{dt}{2\pi} 
\frac{t}{\sqrt{t^2-m_0^2}}\,\left[\nu(t)-\nu^{(1)}(t)\right]
\qquad {\rm where}\qquad
m_0={\rm min}(\mu,2)\,.
\label{eq:Evac} \end{equation}
The subtraction of the Born approximation
\begin{equation}
\nu^{(1)}(t)=\int_{0}^\infty dx\, \left[\frac{V_{11}(x)}{t_1}
+\frac{V_{22}(x)}{t_2}\right]\,,
\label{eq:Born1}\end{equation}
where $V_{ij}$ refers to the elements of the potential matrix,
enforces the no-tadpole renormalization condition.

\section{Numerical results}

With the formalism to compute the VPE of the two-field soliton now fully established, 
we list numerical results in Table~\ref{tab:vpe4} as a function of the model parameter 
$\mu$ and the variational parameter $a$.  For $\mu=2$, there is no variation with $a$, 
because the $\mu=2$ model is merely a duplication of the kink model. Hence this 
case serves as a check of our numerical procedure, reproducing the known result
$E_{\rm vac}=1/\sqrt{3}-6/\pi\approx-1.3325$ \cite{Dashen:1974ci,Ra82}. Our numerical 
results are also consistent with those from a heat kernel expansion. Though we observe 
differences of up to one percent compared to the earlier results~\cite{AlonsoIzquierdo:2003gh}, 
a subgroup of those authors has later modified the expansion for 
$1.4\le\mu\le3.1$ in Refs.~\cite{AlonsoIzquierdo:2012tw,Alonso-Izquierdo:2013pma} and we agree 
with the data reported in Table 1 (with $\gamma=a$ and $\sigma=\mu$) of 
Ref.~\cite{Alonso-Izquierdo:2013pma} to numerical precision.  We also observe a shallow 
local minimum of the VPE for $2<\mu\le3.2$ and moderate $a$. However, this feature is 
not general for $\mu\ge2$. Rather the minimum becomes shallower and its position moves to 
smaller $a$ as $\mu$ increases and fully disappears for $\mu=4.4$. The results produced in 
Tables 4 and 5 of Ref.~\cite{AlonsoIzquierdo:2012tw} suggest that the truncated heat kernel 
expansion is inaccurate for $\mu\le1$. Our spectral methods do not require any truncation and, 
in contrast to statements in Refs.~\cite{AlonsoIzquierdo:2003gh,Alonso-Izquierdo:2013pma},
they clearly show that indeed the DHN method \cite{Dashen:1974ci} (generalizing scattering 
techniques) can be efficiently utilized to compute the VPE of the two-field soliton.
\begin{table}[htbp]
\centerline{
\begin{tabular}{l|cccccccccccc}
\diagbox[height=0.8cm]{~~$a$}{$\mu$} & 0.5  & 0.8  & 1.0  & 1.2  & 1.6
& 2.0 & 2.4 & 2.8 & 3.2 & 3.6 & 4.0 & 4.4 \cr
\hline
0.0     & -0.830 & -0.922 & -0.985 & -1.049 & -1.186 & -1.333 & -1.491 & -1.661 & -1.844 & -2.039 & -2.246 & -2.467\cr
0.1     & -0.833 & -0.924 & -0.986 & -1.050 & -1.186 & -1.333 & -1.491 & -1.661 & -1.843 & -2.038 & -2.246 & -2.467\cr
0.2     & -0.841 & -0.929 & -0.990 & -1.053 & -1.187 & -1.333 & -1.490 & -1.660 & -1.842 & -2.037 & -2.245 & -2.467\cr
0.3     & -0.856 & -0.938 & -0.997 & -1.058 & -1.189 & -1.333 & -1.489 & -1.658 & -1.840 & -2.036 & -2.245 & -2.467\cr
0.4     & -0.878 & -0.952 & -1.007 & -1.065 & -1.192 & -1.333 & -1.487 & -1.656 & -1.839 & -2.035 & -2.247 & -2.470\cr
0.5     & -0.906 & -0.971 & -1.020 & -1.074 & -1.195 & -1.333 & -1.486 & -1.654 & -1.837 & -2.036 & -2.249 & -2.477\cr
0.6     & -0.949 & -0.995 & -1.037 & -1.085 & -1.199 & -1.333 & -1.484 & -1.653 & -1.838 & -2.039 & -2.257 & -2.490\cr
0.7     & -1.007 & -1.027 & -1.059 & -1.100 & -1.204 & -1.333 & -1.483 & -1.653 & -1.841 & -2.048 & -2.273 & -2.515\cr
0.8     & -1.089 & -1.072 & -1.089 & -1.119 & -1.210 & -1.333 & -1.482 & -1.656 & -1.851 & -2.068 & -2.304 & -2.560\cr
0.9     & -1.229 & -1.145 & -1.135 & -1.147 & -1.217 & -1.333 & -1.484 & -1.666 & -1.876 & -2.112 & -2.372 & -2.656\cr
0.99    & -1.661 & -1.363 & -1.271 & -1.227 & -1.235 & -1.333 & -1.496 & -1.714 & -1.978 & -2.284 & -2.628 & -3.008\cr
0.999   & -2.076 & -1.572 & -1.401 & -1.303 & -1.251 & -1.333 & -1.510 & -1.764 & -2.083 & -2.459 & -2.888 & -3.364\cr
0.9999  & -2.488 & -1.781 & -1.530 & -1.379 & -1.268 & -1.333 & -1.523 & -1.813 & -2.187 & -2.634 & -3.147 & -3.720\cr
0.99999 & -2.900 & -1.990 & -1.660 & -1.454 & -1.284 & -1.333 & -1.536 & -1.863 & -2.291 & -2.809 & -3.406 & -4.076
\end{tabular}}
\caption{\label{tab:vpe4}Numerical results for the vacuum polarization energy, Eq.~(\ref{eq:Evac}),
of the Shifman-Voloshin soliton. The $a=0$ results are from I) in Table~\ref{tab:known}. 
The $\mu=0.5$ results have been obtained from both the
analytical and numerical soliton solutions.}
\end{table}

For large enough $a$, curve-fitting suggests a logarithmic divergence as $a\to1$,
\begin{equation}
E_{\rm vac}\sim E_0 + E_1{\rm ln}(1-a)\,,
\label{eq:logfit}\end{equation}
where the constants $E_0$ and $E_1$ vary with $\mu$.  We present our results for $E_1$ 
in Table~\ref{tab:barr}. This logarithmic divergence signals
the instability of the soliton. Regardless of the coupling constant $\lambda$, there is 
always a value of $a$ close enough to one such that the total energy is negative. Again, it is 
important to stress that $a$ is a variational parameter of the soliton rather than a (fixed)
model parameter like $\lambda$ or $\mu$.

To understand the origin of the instability, we note that in the limit $a\to1$, for a wide 
region of space ($|x|\le x_0$), the profiles equal their $x=0$ values. This actually is the 
secondary vacuum at $\phi=0$ and $\chi=\sqrt{2/\mu}$ whose curvature differs from the primary
vacuum at $\phi=1$ and $\chi=0$. In that range, the potential matrix Eq.~(\ref{eq:vpot}) 
can therefore be approximated by a (double) step function
\begin{equation}
V(x)\sim \left(\mu-2\right)
\begin{pmatrix}-\mu & 0 \cr 0 & 2\end{pmatrix} \,\Theta(x_0-|x|)\,.
\label{eq:vpotlim}\end{equation}
In this limit the two channels decouple and the vacuum polarization energy can be 
computed separately. Moreover, this limiting potential is symmetric under $x\to-x$ and
standard techniques yield \cite{Graham:2002xq}
\begin{equation}
E_{\rm B}=\sum_{i=1,2}\int_{m_i}^\infty \frac{dt}{2\pi}\,
\frac{t}{\sqrt{t^2-m_i^2}}\,\left[
{\rm ln}\left\{g_i(t,0)\left(g_i(t,0)-\frac{1}{t}g_i^\prime(t,0)\right)\right\}
\right]_1\,,
\label{eq:EvacJost}
\end{equation}
where $i=1,2$ refers to the two (decoupled) particles with $m_1=\mu$ and $m_2=2$. 
The subscript on the square brackets denotes the subtraction of the Born approximation 
in the no-tadpole renormalization condition. Since the potential matrix is diagonal,
$V={\rm diag}(V_{11},V_{22})$, the Jost solutions $g_i(t,x)$ decouple and obey 
the differential equations
\begin{equation}
g_i^{\prime\prime}(t,x)=2tg_i^\prime(t,x)+V_{ii}(x)g_i(t,x)
\label{eq:DEQJost}
\end{equation}
subject to the boundary conditions $g_i(t,\infty)=1$ and $g_i^\prime(t,\infty)=0$.
For the step-function potential of Eq.~(\ref{eq:vpotlim}) the Jost functions are 
straightforwardly obtained to be \cite{Weigel:2016zbs}
\begin{equation}
g_i(t,0)=\frac{\kappa^{(i)}_1{\rm e}^{-\kappa^{(i)}_2x_0}
-\kappa_2{\rm e}^{-\kappa^{(i)}_1x_0}}{\kappa^{(i)}_1-\kappa^{(i)}_2}
\qquad{\rm and}\qquad
g_i^\prime(t,0)=\frac{\kappa^{(i)}_1\kappa^{(i)}_2}{\kappa^{(i)}_1-\kappa^{(i)}_2}
\left({\rm e}^{-\kappa^{(i)}_2x_0}-{\rm e}^{-\kappa^{(i)}_1x_0}\right)\,,
\label{eq:barrSol}
\end{equation}
with $\kappa^{(i)}_{1,2}=t\pm\sqrt{t^2+v_{i}}$, $v_{1}=\mu(2-\mu)$ and $v_{2}=2(\mu-2)$. 
For the kinematics required by the integral Eq.~(\ref{eq:EvacJost}), $t\ge\mu$ for $i=1$ and 
$t\ge2$ for $i=2$, the $\kappa^{(i)}_{1,2}$ are always real. This suggests that the limiting 
potential solely acts as a barrier for the purpose of the VPE, even though one of the diagonal 
components of $V$ is attractive. We thus expect that the VPE becomes even more negative as we increase 
$x_0$ \cite{Weigel:2016zbs}.
\begin{table}
\centerline{
\begin{tabular}{c|cccccccccccc}
$\mu$ & 0.5  & 0.8  & 1.0  & 1.2  & 1.6
& 2.0 & 2.4 & 2.8 & 3.2 & 3.6 & 4.0 & 4.4 \cr
\hline
$E_1$ & 0.182 & 0.092 & 0.057 & 0.034 & 0.007 & 0 & 0.006 & 0.021 & 0.045 & 
0.075 & 0.112 & 0.154  \cr
$E_{\rm B}/x_0$ & -0.178 & -0.114 & -0.079 & -0.051 & -0.013 &
0.0  & -0.013 & -0.051 & -0.115 & -0.204 & -0.319 & -0.459
\end{tabular}}
\caption{\label{tab:barr}Coefficient of the logarithmic divergence in Eq.~(\ref{eq:logfit})
fitted from $a\ge0.8$ data in Table~\ref{tab:vpe4}, and energy per unit length for step function 
potential, Eq.~(\ref{eq:vpotlim}).}
\end{table}
Indeed numerically we find that for large $x_0$ the ratio $E_B/x_0$ approaches negative 
constants (varying with $\mu$), which we also list in Table~\ref{tab:barr}. Of course,
for large $x_0$ the ratio $E_B/x_0$ is nothing but the quantum energy density of the 
secondary vacuum. We have computed the correlation coefficient between $E_1$ and 
$E_{\rm B}/x_0$ to be 0.8, {\it i.e.} quite significant. This is a first indication 
that the instability is caused by the solution spreading into the secondary vacuum. 
To further compare with the step function potential, Eq.~(\ref{eq:vpotlim}),
we first estimate the extension of the soliton as
\begin{equation}
x_S(a)=\frac{\int_0^\infty dx\, x \epsilon(x)}{\int_0^\infty dx\, \epsilon(x)}
=\frac{3}{2}\int_0^\infty dx\, x \epsilon(x)\,,
\label{eq:exten}\end{equation}
where $\epsilon(x)$ is the energy density of the soliton, {\it i.e.} the
integrand of the first integral in Eq.~(\ref{eq:bps}) with the soliton profiles
for given value of $a$ substituted. From this we obtain
\begin{equation}
\Delta E_{\rm B}=E_{\rm B}(x_S(0.99))-E_{\rm B}(x_S(0.99999))\,.
\label{eq:delEB}\end{equation}
In Table~\ref{tab:comp} we compare $\Delta E_{\rm B}$ with $\Delta E_{\rm S}$, the latter 
being the difference between the $a=0.99$ and $a=0.99999$ entries in Table~\ref{tab:vpe4}. 
\begin{table}[t]
\centerline{
\begin{tabular}{c|ccccccccccc}
$\mu$ & 0.5  & 0.8  & 1.0  & 1.2  & 1.6
& 2.4 & 2.8 & 3.2 & 3.6 & 4.0 & 4.4 \cr
\hline
$\Delta E_{\rm B}$ & 1.242 & 0.627 & 0.390 & 0.228 & 0.049 & 0.040 &
0.149 & 0.313 & 0.527 & 0.779 & 1.072\cr
$\Delta E_{\rm S}$ & 1.239 & 0.627 & 0.390 & 0.227 & 0.049 & 0.040 &
0.149 & 0.313 & 0.525 & 0.777 & 1.068
\end{tabular}}
\caption{\label{tab:comp}Differences of vacuum polarization energies for soliton extensions
(into the secondary vacuum) corresponding to $a=0.99$ and $a=0.99999$. Top row: limiting potential,
Eq.~(\ref{eq:vpotlim}), bottom row: full potential with data taken from Table~\ref{tab:vpe4}.}
\end{table}
The coincidence (to numerical accuracy) between  $\Delta E_{\rm B}$ and $\Delta E_{\rm S}$ 
clearly shows that the instability is caused by the expansion of the soliton into the secondary vacuum.

One might imagine that finite renormalizations required to implement an on-shell 
scheme could alter this conclusion. In the appendix we determine the counterterms for
this scheme and show that the instability persists.

\section{Conclusion}

For a well-established soliton model consisting of two real scalar fields in one space 
dimension, we have computed the one-loop quantum correction to the soliton energy for a 
complete set of soliton solutions. The model is defined in terms of three parameters, the 
coupling strength $\lambda$ and the masses of the two fields. For simplicity we have 
scaled coordinates and fields such that $\lambda$ only appears as an overall factor 
in the action but not in the field equations, which are only sensitive to the mass 
ratio. However, upon quantization the classical energies carry a relative factor of 
$1/\lambda$ compared to the quantum energies. In this model, the soliton solutions 
are characterized by two variational parameters, the location of the soliton and the 
amplitude $a<1$ of the spatially symmetric profile function. While the classical energy 
does not vary with these parameters, the vacuum polarization energy diverges 
logarithmically to negative infinity as $a\to1$ unless the two masses are equal. We 
have employed (generalized) spectral methods \cite{Weigel:2017kgy,Graham:2009zz} for 
this computation but note that our results are consistent with those obtained in a heat 
kernel expansion \cite{AlonsoIzquierdo:2003gh,AlonsoIzquierdo:2012tw,Alonso-Izquierdo:2013pma}. 
However, those studies did not identify the importance and origin of this divergence,\footnote{In 
Ref. \cite{AlonsoIzquierdo:2003gh} it was concluded that the VPE would be almost 
constant for a wide range of $a$. This is indeed true for $\mu \approx 2$ and $a$
not too close to one, but is not the case otherwise.} which implies that 
for any finite value of $\lambda$, we can find $a\lesssim1$ such that the total energy 
is negative, and hence the quantum corrections destabilize the solitons. We have seen 
that the instability is related to the existence of a {\it secondary} vacuum. While 
this configuration is classically degenerate with the {\it primary} vacuum, its 
curvatures in field  space, {\it i.e.} the masses of the fluctuating fields, are 
different, as given by the matrix in Eq.~(\ref{eq:vpotlim}). As $a\to1$, the soliton 
profiles contain larger and larger regions of the secondary vacuum, thereby reducing
the vacuum polarization energy. We have confirmed this feature by comparison with the 
vacuum polarization energy of a step function background simulating the transition
between the two vacua.

The situation is similar to the $\varphi^6$ model with
$U_6=\fract{1}{2}\left(\varphi^2+\alpha^2\right)\left(\varphi^2-1\right)^2$;
in that case the range of the secondary vacuum is measured by the 
model parameter $\alpha$ and the vacuum polarization energy diverges as $\alpha\to0$ 
\cite{AlonsoIzquierdo:2011dy,Weigel:2016zbs}. In that model the primary vacua are at 
$\varphi=\pm1$ and soliton solutions connect these vacua between negative and positive 
spatial infinity. As $\alpha\to0$ a secondary vacuum emerges at $\varphi=0$ and the 
soliton assumes that value\footnote{For $\alpha=0$ two solitons exist~\cite{Lohe:1979mh}; 
one connects $\phi=-1$ to $\phi=0$ and the other $\phi=0$ to $\phi=1$. The $\alpha\ne0$ 
version combines these transitions. As $\alpha$ increases, their overlap within the secondary
vacuum decreases.} in a range of space that increases with $1/\alpha$. Together with the 
current study this suggests that quantum instabilities of solitons emerge as a generic 
feature whenever there is a secondary vacuum.

As a next step it will be interesting to see whether this instability persists when
embedding this one space dimension model into higher dimensions, {\it i.e.} whether domain 
walls constructed from this model will also be unstable. This investigation can be pursued 
along the line of the interface formalism \cite{Graham:2001dy}. With additional 
quantum fluctuations in the transverse directions, the vacuum polarization energy per 
unit length/area will be altered, while the classical energy (also per unit length/area) 
will remain unchanged because it is a local quantity. 

\acknowledgments
The authors very much appreciate helpful discussions with M. Quandt at an early stage
of this project.
H.~W.\ is supported in part by the National Research Foundation of South Africa (NRF)
by grant~109497. N.~G.\ is supported in part by the National Science Foundation (NSF)
through grant PHY-1520293.

\appendix

\section{Effective action and renormalization}

We determine the counterterms for on-shell renormalization from the one-loop effective action
\begin{equation}
\mathcal{A}_{\rm eff}=\frac{\imu}{2}{\rm Tr}\,{\rm ln}\left[1+G^{-1}V\right]
\qquad {\rm where}\qquad
G=\partial_\mu\partial^\mu+M^2
\label{eq:aeff}\end{equation}
is the mass matrix defined in Eq.~(\ref{eq:master}) and $V$ is the potential 
matrix from Eq.~(\ref{eq:vpot}). The trace in Eq.~(\ref{eq:aeff}) goes over 
the two particle species and space-time. Only the contribution linear in $V$ 
is ultra-violet divergent. In $D=2-2\epsilon$ space-time dimensions,\footnote{The
equivalence with the Born result, Eq.~(\ref{eq:Born1}) has been established in 
Ref.~\cite{Farhi:2000ws} for arbitrary dimensions.} this 
linear term becomes
\begin{align}
\mathcal{A}^{(1)}_{\rm eff}&=
-\frac{1}{8\pi}\left[\frac{1}{\epsilon}-\gamma
-{\rm ln}\frac{\mu^2}{4\pi^2\Lambda^2}\right]\int d^2x\, V_{11}
-\frac{1}{8\pi}\left[\frac{1}{\epsilon}
-\gamma-{\rm ln}\frac{4}{4\pi^2\Lambda^2}\right]\int d^2x\, V_{22}\,,
\label{eq:aeff1}\end{align}
where $\Lambda$ is a renormalization scale (which is dimensionless in the present 
conventions). In this divergent term the fields appear at most quadratically and we 
need the two counterterms
\begin{equation}
\mathcal{L}^{(1)}_{\rm ct.}=c_1\left(\phi^2-1\right)+c_2\chi^2\,,
\label{eq:lct1}\end{equation}
with 
\begin{align}
c_1&=\frac{1}{8\pi}\left\{\mu(\mu+1)\left[\frac{1}{\epsilon}-\gamma
-{\rm ln}\frac{\mu^2}{4\pi^2\Lambda^2}\right]
+6\left[\frac{1}{\epsilon}-\gamma-{\rm ln}\frac{4}{4\pi^2\Lambda^2}
\right]\right\}\cr
c_2&=\frac{1}{8\pi}\left\{\frac{3}{2}\mu^2\left[\frac{1}{\epsilon}-\gamma
-{\rm ln}\frac{\mu^2}{4\pi^2\Lambda^2}\right]
+\mu(\mu+1)\left[\frac{1}{\epsilon}-\gamma-{\rm ln}\frac{4}{4\pi^2\Lambda^2}
\right]\right\}\,.
\label{eq:c1c2}\end{align}
Note that these counterterms do not form part of the original Lagrangian,
Eq.~(\ref{eq:Bazeia}), preventing a direct multiplicative renormalization. However,
the only purpose of this counterterm is to exactly compensate the local tadpole 
diagram contribution to the action, Eq.~(\ref{eq:aeff1}), which is represented 
by the Born approximation, Eq.~(\ref{eq:Born1}), in the VPE. Hence this counterterm
will cancel in the total energy.

To implement the physical on-shell scheme we define fluctuations $\eta_i$ about 
the primary vacuum via $\phi=1+\eta_1$ and $\chi=0+\eta_2$. The counterterms
\begin{align}
\mathcal{L}^{(2)}_{\rm ct.}&=-d_1\left(\phi^2-1+\frac{\mu}{2}\chi^2\right)^2
-d_2\left(\chi\phi\right)^2
+d_3\partial_\mu\phi\partial^\mu\phi+d_4\partial_\mu\chi\partial^\mu\chi\cr
&=-4d_1\eta_1^2-d_2\eta_2^2+d_3\partial_\mu\eta_1\partial^\mu\eta_1
+d_4\partial_\mu\eta_2\partial^\mu\eta_2+\mathcal{O}(\eta_i^3)\,.
\label{eq:lct2}\end{align}
compensate for any quantum correction to the positions and residues of the
$\eta_i$ propagators. We stress that $\mathcal{L}^{(2)}_{\rm ct.}$ contains 
exactly the terms of the original Lagrangian, Eq.~(\ref{eq:Bazeia}). That is, 
the on-shell scheme can indeed be implemented by multiplicative renormalization, 
at least at one loop order. This counterterm will contribute
\begin{equation}
E_{\rm ct.}=d_1\int dx\,
\left(\phi^2-1+\frac{\mu}{2}\chi^2\right)^2
+\frac{d_2}{\mu^2}\int dx\, \left(\mu\chi\phi\right)^2
+d_3\int dx\, \phi^{\prime2} +d_4\int dx\, \chi^{\prime2}
\label{eq:ect}\end{equation}
to the total energy. When substituting the soliton configuration, Derrick's theorem
tells us that
$$
\frac{1}{2}\int dx\, \left[\left(\phi^2-1+\frac{\mu}{2}\chi^2\right)^2
+\left(\mu\chi\phi\right)^2\right]=
\frac{1}{2}\int dx\, \left[\phi^{\prime2}+\chi^{\prime2}\right]=\frac{2}{3}\,.
$$
Hence each of the integrals in Eq.~(\ref{eq:ect}) is bounded between zero and $\frac{4}{3}$.
By renormalizability, the coefficients $d_i$ do not depend on the profiles. Hence 
$E_{\rm ct.}$ will only contribute a finite amount to the total energy in the limit
$a\to1$ and thus its addition does not prevent the instability.

For completeness we determine the finite coefficients $d_i$ from the second order effective 
action $\mathcal{A}^{(2)}_{\rm eff}=-\frac{\imu}{4}{\rm Tr}\,\left[ G^{-1}VG^{-1}V\right]$.
To compare with the expanded form in Eq.~(\ref{eq:lct2}) it suffices to consider
the linear terms in~$V$: 
$$
V=\begin{pmatrix}2\mu(\mu+1)\eta_1 & 2\mu(\mu+1)\eta_2 \cr
2\mu(\mu+1)\eta_2 & 12 \eta_1\end{pmatrix}+\mathcal{O}(\eta_i^2)\,.
$$
Performing the loop integrals yields
\begin{align}
\mathcal{A}^{(2)}_{\rm eff}&=\frac{1}{4\pi}\int \frac{d^2k}{2\pi}\,
\left\{\left[\mu^2(\mu+1)^2I(k^2,2,2)+36I(k^2,\mu,\mu)\right]
\widetilde{\eta}_1(k)\widetilde{\eta}_1(-k)
+2\mu^2(\mu+1)^2I(k^2,2,\mu)\widetilde{\eta}_2(k)\widetilde{\eta}_2(-k)
\right\}
\label{eq:aeff2b}\end{align}
where $\widetilde{\eta}_i(k)=\int d^2x\, {\rm e}^{ik\cdot x}\eta_i(x)$ are Fourier
transforms and
$$
I(k^2,m_1,m_2)=\int_0^1\, \frac{dx}{m_1^2-x(m_1^2-m_2^2)-x(1-x)k^2}
$$
is a Feynman parameter integral. The gradient expansion is the Taylor expansion in $k^2$
$$
I(k^2,m_1,m_2)=I_1(m_1,m_2)+I_2(m_1,m_2)k^2+\ldots
$$
with 
\begin{align}
I_1(m_1,m_2)&=I(0,m_1,m_2)=\frac{{\rm ln}\left(\frac{m_1^2}{m_2^2}\right)}{m_1^2-m_2^2}
=\frac{{\rm ln}\left(\frac{m_2^2}{m_1^2}\right)}{m_2^2-m_1^2}\,,\cr
I_2(m_1,m_2)&=\frac{\partial}{\partial (k^2)}I(k^2,m_1,m_2)\Big|_{k^2=0}
=\frac{1}{(m_2^2-m_1^2)^3}\left[2m_1^2-2m_2^2
+(m_1^2+m_2^2){\rm ln}\left(\frac{m_2^2}{m_1^2}\right)\right]\,,
\nonumber\end{align}
when $m_1\ne\ m_2$ while
$I_1(m,m)=1/m^2$ and $I_2(m,m)=1/(6m^4)$.
With the gradient expansion we return to coordinate space
\begin{align}
\mathcal{A}^{(2)}_{\rm eff}=\frac{1}{4\pi}\int d^2x\,\Big\{&
\left[\mu^2(\mu+1)^2I_1(2,2)+36I_1(\mu,\mu)\right]\eta_1^2
+2\mu^2(\mu+1)^2I_1(2,\mu)\eta_2^2\cr
&+\left[\mu^2(\mu+1)^2I_2(2,2)+36I_2(\mu,\mu)\right]
\partial_\mu\eta_1\partial^\mu\eta_1
+2\mu^2(\mu+1)^2I_2(2,\mu)\partial_\mu\eta_2\partial^\mu\eta_2\Big\}+\ldots\,.
\label{eq:aeff2c}\end{align}
In the on-shell renormalization the quadratic terms in
$\mathcal{A}^{(2)}_{\rm eff}$ are canceled by those in $\mathcal{L}_{\rm ct.}$:
\begin{equation}
\begin{array}{llll}
d_1&=\frac{1}{4}\left[\mu^2(\mu+1)^2I_1(2,2)+36I_1(\mu,\mu)\right]\hspace{2cm}
&d_2&=2\mu^2(\mu+1)^2I_1(2,\mu)\cr
d_3&=-\mu^2(\mu+1)^2I_2(2,2)+36I_2(\mu,\mu)
&d_4&=-2\mu^2(\mu+1)^2I_2(2,\mu)\,.
\end{array}
\label{eq:rencoef}\end{equation}

\end{document}